\newcommand{\beq}{\begin{equation}}
\newcommand{\eeq}{\end{equation}}
\newcommand{\bea}{\begin{eqnarray}}
\newcommand{\eea}{\end{eqnarray}}

\newcommand{\gsim}{\lower.7ex\hbox{$\;\stackrel{\textstyle>}{\sim}\;$}}
\newcommand{\lsim}{\lower.7ex\hbox{$\;\stackrel{\textstyle<}{\sim}\;$}}

\documentclass[aps,prev,twocolumn,preprintnumbers,floatfix,nofootinbib]{revtex4-1}
\usepackage{graphicx}
\usepackage{epstopdf}
\usepackage{mathrsfs}
\usepackage{amssymb}
\usepackage{verbatim}
\usepackage{color}
\usepackage{multirow}
\def\stacksymbols #1#2#3#4{\def\theguybelow{#2}
    \def\vp{\lower#3pt}
    \def\sp{\baselineskip0pt\lineskip#4pt}
    \mathrel{\mathpalette\intermediary#1}}

\def\intermediary#1#2{\vp\vbox{\sp
     \everycr={}\tabskip0pt
     \halign{$\mathsurround0pt#1\hfil##\hfil$\crcr#2\crcr
              \theguybelow\crcr}}}

\def\be{\begin{equation}}
\def\ee{\end{equation}}
\def\bea{\begin{eqnarray}}
\def\eea{\end{eqnarray}}

\def\sp{\;\;\;,\;\;\;}

\def\lsim{\raise0.3ex\hbox{$\;<$\kern-0.75em\raise-1.1ex\hbox{$\sim\;$}}}
\def\gsim{\raise0.3ex\hbox{$\;>$\kern-0.75em\raise-1.1ex\hbox{$\sim\;$}}}

\def\inbar{\,\vrule height1.5ex width.4pt depth0pt}

\def\IC{\relax\hbox{$\inbar\kern-.3em{\rm C}$}}
\def\IQ{\relax\hbox{$\inbar\kern-.3em{\rm Q}$}}
\def\IR{\relax{\rm I\kern-.18em R}}
 \font\cmss=cmss10 \font\cmsss=cmss10 at 7pt
\def\IZ{\relax\ifmmode\mathchoice
 {\hbox{\cmss Z\kern-.4em Z}}{\hbox{\cmss Z\kern-.4em Z}}
 {\lower.9pt\hbox{\cmsss Z\kern-.4em Z}}
 {\lower1.2pt\hbox{\cmsss Z\kern-.4em Z}}\else{\cmss Z\kern-.4em Z}\fi}

\def\comment#1{}

\def\u1x{U(1)_X}
\newcommand{\nc}{\newcommand}
\nc{\LL}{L}
\nc{\vv}{\tilde{v}}
\nc{\ccdot}{\!\cdot\!}
\nc{\gsm}{G_{SM}}
\nc{\vfive}{\mathbf{5}\oplus\mathbf{\overline{5}}}
\nc{\vten}{\mathbf{10}\oplus\mathbf{\overline{10}}}
\nc{\zhol}{Z^{\rm hol}}
\nc{\xfb}{\,{\rm fb}}

\begin{document}

%\wideabs{
%\begin{flushright}
%
%
%\end{flushright}

\preprint{LPT--Orsay 13-12}
\preprint{IFT-UAM/CSIC-13-014}
\preprint{UMN--TH--3137/13}
\preprint{FTPI--MINN--13/04}

\vspace*{1mm}

\title{Gauge Coupling Unification and Non-Equilibrium Thermal Dark Matter}

\author{Yann Mambrini$^{a}$}
\email{yann.mambrini@th.u-psud.fr}
\author{Keith A. Olive$^{b}$}
\email{olive@physics.umn.edu}
\author{J\'er\'emie Quevillon$^{a}$}
\email{jeremie.quevillon@th.u-psud.fr}
\author{Bryan Zald\'\i var$^{c}$}
\email{bryan.zaldivar@uam.es}

\vspace{0.1cm}
\affiliation{
${}^a$ Laboratoire de Physique Th\'eorique 
Universit\'e Paris-Sud, F-91405 Orsay, France.
 }
 \affiliation{
${}^b$ 
 William I.~Fine Theoretical Physics Institute, 
       School of Physics and Astronomy,
            University of Minnesota, Minneapolis, MN 55455, USA
}
 
\affiliation{
${}^c$ 
Instituto de Fisica Teorica, IFT-UAM/CSIC, 
 28049 Madrid, Spain 
}

\begin{abstract} 

We study a new mechanism for the production of dark matter in the universe
which does not rely on thermal equilibrium. Dark matter is populated 
from the thermal bath subsequent to inflationary reheating via a massive mediator whose mass is above the reheating scale, $T_{RH}$. To this end,
we consider models with an extra U(1) gauge symmetry broken at some
intermediate scale ($M_{int}\simeq 10^{10-12}$GeV).
We show that not only does the model allow for gauge coupling unification
(at a higher scale associated with grand unification) but can naturally provide 
a dark matter candidate which is a Standard Model singlet but charged under the extra U(1).
The intermediate scale gauge boson(s) which are
predicted in several E6/SO(10) constructions can be a natural mediator between dark matter 
and the thermal bath. We show that the dark matter abundance, while never having achieved
thermal equilibrium, is fixed shortly after the reheating epoch by the relation
$T_{RH}^3/M_{int}^4$. 
 As a consequence, we show that the unification of gauge couplings which determines $M_{int}$
 also fixes
 the reheating temperature, which can be as high as  
  $T_{RH}\simeq10^{11}$GeV.
  
\end{abstract}

\maketitle
%}

%%%%%%%%%%%%%%%%%%%%%%%%%%%%%%%%%%%%%%%%%%%%%%%%%%%%%%%%%%%%%%%%%
%%%%%%%%%%%%%%%%%%%%%%%%%%%%%%%%%%%%%%%%%%%%%%%%%%%%%%%%%%%%%%%%%
%%%%%%%%%%%%%%%%%%%%%%%%%%%%%%%%%%%%%%%%%%%%%%%%%%%%%%%%%%%%%%%%%

\maketitle

%% The arXiv's use of hypertex conflicts with revtex4's use of
%% \tableofcontents in single column format. To avoid this problem,
%% Include a file OOREADME.xxx with the word nohypertex in it when
%% you submit to the arXiv.
%\tableofcontents

%\section{Introduction}\label{sec:introduction}
\setcounter{equation}{0}

%%%%%%%%%

%%%%%%%%%%%%%%%%%%%%%%%%%%%%%%%%%%%%%%%%%%%%%%%%%%%%%%%%%%%%%%%%%%%%%%

\section{Introduction}

The Standard Model (SM) of particle physics is now more than ever motivated by the 
recent discovery of the Higgs boson at both the ATLAS \cite{HiggsATLAS} and CMS
\cite{HiggsCMS} detectors. The SM, however,  contains many free parameters, and the gauge couplings do not unify.  Among the most elegant approaches to understand some of these parameters is the idea
of a grand unified theory (GUT) in which the three gauge couplings $\alpha_{1,2,3}$ 
originate from a single gauge coupling associated to a 
grand unified gauge group \cite{GUT}. This idea is supported by the fact that quantum numbers
of quarks and leptons in the SM nicely fill representations of a GUT symmetry, e.g., the
{\bf 10} and $\bar {\bf 5}$ of $SU(5)$ or {\bf 16} of $SO(10)$. 

Another issue concerning the SM is the lack of a candidate to account
for Dark Matter (DM) which consists of 22 \% of the energy density of our universe.
Stable Weakly Interacting Massive Particles (WIMPs) are among the most popular candidates for DM.
In most models, such as popular supersymmetric extensions of the SM \cite{EHNOS}, the annihilation
of WIMPs in thermal equilibrium in the early universe determined the relic abundance of DM. 

In this letter, we will show that GUT gauge groups such as E$_6$ or SO(10)
which contain additional U(1) gauge subgroups and are broken at
an intermediate scale, can easily lead to gauge coupling unification \cite{Fukugita:1993fr}
and may contain a new dark matter candidate which is charged under the 
extra U(1). However, unlike the standard equilibrium annihilation process, or complimentary
process of freeze-in \cite{Freezein}, we propose an alternative mechanism
for producing dark matter through interactions which are mediated by
the heavy gauge bosons associated with the extra U(1). While being produced from
the thermal bath, these dark matter particles never reach equilibrium. We will refer
to dark matter produced with this mechanism as Non-Equilibrium Thermal Dark Matter or NETDM.
The final relic abundance of NETDM is obtained shortly after the inflationary reheating epoch. This mechanism is fundamentally different from other non-thermal DM production 
mechanisms in the literature (to our knowledge). 
Assuming that none of the dark matter particles are directly produced by
the decays of the inflaton during reheating, we compute the production of 
dark matter and relate the inflationary reheat temperature to the choice of 
the gauge group and the intermediate scale needed
for gauge coupling unification. As an added benefit, the model 
naturally possesses the capability of producing a baryon asymmetry through leptogenesis, although that lies beyond of the scope of this work.

 \noindent
 The letter is organized as follows. After a summary of the unified models under consideration 
 in section II, 
 we show how the presence of an intermediate scale allows for the possibility of producing 
 a dark matter candidate which respects the WMAP constraint \cite{WMAP}
and apply it to an explicit scenario in section III. A discussion of our main result is found
in section IV before  concluding in section V.

\section{Unification in SO(10) models}
\label{sec:unification}

The prototype of grand unification is based on the SU(5) gauge group. In an extension of SU(5)
one can introduce SU(5) singlets as potential dark matter candidates. The simplest extension
in which singlets are automatically incorporated is that of SO(10). 
There are, however, many ways to break
SO(10) down to $SU(3)\times SU(2)\times U(1)$. 
This may happen in multiple stages, but here we are
mainly concerned with the breaking of an additional
U(1) (or SU(2)) factor at an intermediate scale $M_{int}$.
Here, we will not go into the details of the breaking, 
but take some specific, well-known examples when needed. 
Assuming gauge coupling unification, the GUT mass scale, $M_{GUT}$, 
and the intermediate scale $M_{int}$ 
can be predicted from the low--energy
coupling constants with the use of the renormalisation group equation,
\be
\mu \frac{d \alpha_i}{d \mu} = -b_i \alpha_i^2~.
\label{rge}
\ee

The evolution of the three running coupling constants $\alpha_1$, $\alpha_2$ and $\alpha_3$
from $M_Z$ to the intermediate scale $M_{int}$ is obtained from Eq.(\ref{rge}) using the
$\beta$--functions of the Standard Model: $b_{1,2,3} = (-41/10, 19/6, 7)/2\pi$.
We note that the gauge coupling, $g_D$, associated with $U'(1)$ is related
at the GUT scale to $g_1$ of $U(1)_Y$ by 
$g_D = \sqrt{\frac{5}{3}}g_1$ and $\alpha_i = g_i^2/4\pi$.    
Between $M_{int}$ and $M_{GUT}$ (both to be determined) the running
coupling constants are again obtained from Eq.(\ref{rge}), now using $\beta$--functions associated
with the intermediate scale gauge group, which we will label $\tilde b_i$.
The matching condition between the two different runnings at $M_{int}$ can be written:
\be
(\alpha_i^0)^{-1} + b_i (t_{int}-t_Z) = \alpha^{-1} + \tilde b_i (t_{int} - t_{GUT})
\label{Eq:matching}
\ee
with $t_{int}= \ln M_{int}$,  $t_Z = \ln M_Z$, 
$t_{GUT}= \ln M_{GUT}$, $\alpha_i^0 = \alpha_i(M_Z)$ which is measured,
and $\alpha = \alpha_i(M_{GUT})$  is the unified coupling constant
at the GUT scale. 
This gives us a system of 3 equations, for 3 unknown parameters: $\alpha, t_{int}, t_{GUT}$.
Solving the Eq.(\ref{Eq:matching}), we obtain
\bea
&&
t_{int} = \frac{1}{b_{32}-b_{21}} \biggl[ \frac{(\alpha_3^0)^{-1} -(\alpha_2^0)^{-1}}{\tilde b_2 - \tilde b_3} 
- \frac{(\alpha_2^0)^{-1} - (\alpha_1^0)^{-1} }{ \tilde b_1 - \tilde b_2 }
\biggr.
\nonumber
\\
&&
+(b_{32} - b_{21} ) t_Z
\biggl.  \biggr] \, ,
\label{Eq:unif}
\eea
where $b_{ij} \equiv (b_i - b_j)/(\tilde b_i - \tilde b_j)$.

\begin{figure}[t]
    \begin{center}
    \includegraphics[width=3.in]{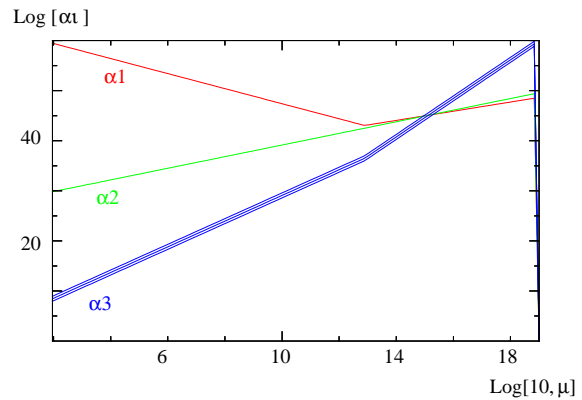}
              \caption{{\footnotesize
             Example of the running of the SM gauge couplings for  $SO(10) \rightarrow SU(4) \times SU(2)_L \times U(1)_R $. 
}}
\label{Fig:unif}
\end{center}
\end{figure}

To be concrete, we will consider a specific example to derive
numerical results for the case of the breaking of $SO(10)$: $
SO(10) \rightarrow SU(4)  \times SU(2)_L \times U(1)_R 
 \rightarrow_{M_{int}} SU(3)_C \times SU(2)_L \times U(1)_Y \rightarrow_{M_{EW}} SU(3)_C \times U(1)_{em}$.
 When the intermediate symmetry is broken by a {\bf 16} of Higgs bosons, the $\tilde b_i$ 
 functions are given by $ \tilde b_{1,2,3} =( 5/2, 19/6, 63/6)/{2 \pi}$ \cite{Fukugita:1993fr}, where the computation was done at 1-loop level. For this case, 
 we obtain $ M_{int} = 7.8 \times 10^{12}$ GeV  and $M_{GUT} = 1.3 \times 10^{15}$ GeV using 
$(\alpha_{1,2,3}^0)^{-1} \simeq (59.47, 29.81, 8.45)$.
The evolution of the gauge couplings for this example is shown in  Fig.~\ref{Fig:unif}.

\section{Heavy Z' and dark matter}

It has been shown in \cite{Kadastik:2009cu} and \cite{Frigerio:2009wf} that a stable
dark matter candidate may arise in SO(10) models  from an unbroken  $Z_2^{B-L}$ symmetry. If 
the dark matter is a fermion (scalar) it should belong to a $3(B-L)$ even (odd) representation of SO(10).
For example, the ${\bf 126}$ or ${\bf 144}$  contains a stable component $\chi$ which is neutral under the SM, 
yet charged under the extra U(1). As we have seen, to explain the unification of the gauge couplings in SO(10)
one needs an intermediate scale $M_{int}$ of order $10^{10}$ GeV. 
The dark matter candidate, $\chi$, can be produced in the early Universe
through s-channel $Z'$  exchange: $SM ~ SM \rightarrow Z'   \rightarrow \chi ~ \chi$.
Since $M_{Z'} = \frac{5}{\sqrt{3}} g_D~ M_{int}$,
the exchanged particle is so heavy (above the reheating scale, as we show below) that the DM production rate
is very slow, and we can neglect the self annihilation process in the Boltzmann equation.
Thus while the dark matter is produced from the thermal bath, we have a non--equilibrium production mechanism for dark matter, hence NETDM.

The evolution of the yield of $\chi$, $Y_\chi = n_{\chi}/{\bf s}$ follows
\be
\frac{dY_\chi}{dx}=\sqrt{\frac{\pi}{45}}\frac{g_s}{\sqrt{g_\rho}} m_\chi M_P \frac{\langle \sigma v \rangle}{x^2} Y_{eq}^2 
\ee
where $n_{\chi}$ is the number density of $\chi$ and
${\bf s}$ the entropy of the universe, 
$g_\rho,g_s$ are the effective degrees of freedom for energy density and entropy, respectively;  $x=m_\chi/T$, $m_\chi$ being the dark matter mass, $M_P$ the Planck mass and
\be
\langle \sigma v \rangle n_{eq}^2 \approx  \frac{\kappa^2~T}{2048 \pi^6} \int_{4 m_\chi^2}^\infty  ds d\Omega \sqrt{s - 4m_\chi^2} |{\cal M}|^2
K_1(\sqrt{s}/T) ~.
\ee
Here $n_{eq}$ is the equilibrium number density of the initial state (SM) particles; and $K_1$ is the first order modified Bessel function
and $\kappa$ the effective degrees
of freedom of incoming particles. 

Since the production of DM occurs mainly at $T_{RH} \gg m_\chi$, we can neglect $m_\chi$ in
estimating the amplitude for production. 
In this case, assuming that both $\chi$ and the initial state, $f$, are fermions, we obtain
\be
\label{ampDMfer}
|{\cal M}_\chi|^2 \approx \frac{g_D^4 q_\chi^2 q_f^2 N^f_c}{(s - M^2_{Z'})^2} \biggl[ s^2(1 + \cos^2 \theta) \biggl]
\ee
where $\theta$ is the angle between the two outgoing DM particles, 
$N_c^f$ is number of colors of the particle $f$, and $q_i$ is the charge of the particle $i$
under $U'(1)$ with a gauge coupling $g_D$. Here, $q$ is an effective coupling which
will ultimately depend on the specific intermediate gauge group chosen. 
With the approximations $m_\chi,m_f \ll \sqrt{s}$ 
and $M_{Z'}\gg T_{RH}$, and  after integration over $\theta$  and sum over all incoming SM fermions in the thermal bath, we obtain
\be
\frac{dY_\chi}{dx} = \sum_f \frac{g_D^4 q_\chi^2 q_f^2 N^f_c}{x^4}\left( \frac{45}{\pi} \right)^{3/2} \frac{1}{g_s \sqrt{g_\rho}} \frac{m_\chi^3 M_P}{M_{Z'}^4}\frac{\kappa_f^2}{2 \pi^7} 
\label{Eq:Boltzmann}
\ee

Solving Eq.(\ref{Eq:Boltzmann}) between the reheating temperature and a temperature $T$ gives
\be
Y_\chi(T)= \sum_f q_\chi^2 q_f^2 N^f_c \left(\frac{45}{g_s \pi} \right)^{3/2} \frac{M_P}{M_{int}^4}\frac{3~\kappa_f^2}{1250 \pi^7} \biggl[ T_{RH}^3 - T^3 \biggr]
\label{Eq:solution}
\ee
where we replaced the mass of the $Z'$ by $M_{Z'}= \frac{5}{\sqrt{3}} g_D M_{int}$ and made the approximation $g_{\rho} = g_s$.
We note that the effect of $Z'$ decay on the abundance of $\chi$ is completely negligible due to its Boltzmann
suppression in  the Universe: the $Z'$ is largely decoupled from the thermal bath already at the time of reheating.

We note several interesting features from Eq.(\ref{Eq:solution}). First of all, the number density of the dark
matter does not depend at all on the strength of the $U'(1)$ coupling $g_D$ but rather on the intermediate scale (that is determined
by requiring gauge coupling unification as we demonstrated in the previous section). 
Second, the production of dark matter is mainly 
achieved at reheating. 
Thirdly, once the relic abundance is obtained, the number density per comoving frame ($Y$) is fixed, never having reached
thermal equilibrium with the bath.
And finally, upon applying the WMAP determination for the DM abundance, we obtain 
a tight constraint on
$T_{RH}$ once the pattern of $SO(10)$ breaking is known (and thus $M_{int}$ fixed). 

\begin{figure}[t]
    \begin{center}
    \includegraphics[width=3.in]{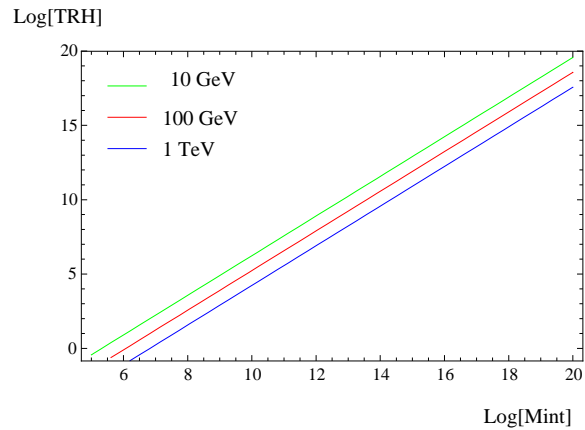}
              \caption{{\footnotesize
          Reheating temperature as function of the SO(10) breaking scale for different mass of dark matter : 10, 100 and 1000 GeV}}
\label{Fig:TRH}
\end{center}
\end{figure}

Thus, 
given a scheme of $SO(10)$ breaking we can determine the reheating temperature very precisely from the relic abundance constraint in the Universe.  From
\be
Y_0 = \frac{\Omega }{m_\chi} \frac{\rho_0^{crit}}{{\bf s}_0} = \left(\frac{\Omega h^2 }{0.1} \right) \frac{13.5}{16 \pi^3} 
\frac{H_0^2 M_P^2}{g_s^0 T_0^3 m_\chi}
\label{Eq:Yinfty}
\ee
where $H$ is the Hubble parameter and the index ``$0$" corresponds to present-day values. Combining Eq.(\ref{Eq:solution}) and Eq.(\ref{Eq:Yinfty})
we find
\be
T^3_{RH} = \frac{5625 ~ \pi^4}{16  q_\chi^2 \sum_f  \kappa^2_f q_f^2 N^f_c} \left( \frac{\Omega h^2 }{0.1} \right) \left( \frac{g_s \pi}{45} \right)^{3/2}
\frac{M_P H_0^2}{T_0^3 m_\chi g_s^0} M_{int}^4
\ee
or 
\be
T_{RH} \simeq 2 \times 10^{8} \mathrm{GeV} \left(  \frac{\Omega h^2 }{0.1}\right)^{1/3} \left( \frac{100\mathrm{GeV}}{m_\chi} \right)^{1/3}
\left( \frac{M_{int}}{10^{12} \mathrm{GeV}} \right)^{4/3}
\ee
where we took for illustration $q_\chi^2 \sum_f \kappa_f^2 q_f^2 N^f_c =1$. 
We show in Fig.(\ref{Fig:TRH}) the evolution of $T_{RH}$ as function  of $M_{int}$  for different values of the dark matter mass $m_\chi$.
We can thus determine the reheating temperature predicted by different symmetry breaking patterns\footnote{We note that
the value obtained for the intermediate scale in different $SO(10)$ breaking schemes is not modified by the presence
of a dark matter particle which is not charged under the SM gauge group.}.  We summarize them in Table \ref{tab:schemes}, 
%%  from bryan
where the values of $T_{RH}$ are given for $m_\chi = 100 $GeV.

%\hspace{-1.cm}
\begin{table}
\caption{\footnotesize{Possible breaking schemes of SO(10).}}
\begin{tabular}{|c|c|c|c|}
\hline
$~$ & $SO(10) \rightarrow {\cal G} \times$  [Higgs]  &$M_{int}$(GeV) & $T_{RH}$(GeV)\\
 \hline
A & $4\times 2_L\times 1_R ~ [{\bf 16}]$ & $10^{12.9}$ & $3 \times 10^9$  \\
 \hline
 A & $ 4\times 2_L\times 1_R ~ [{\bf 126}]$ & $10^{11.8}$ & $1 \times 10^8$  \\
\hline
 B & $4\times 2_L\times 2_R~ [{\bf 16}]$ & $10^{14.4}$ & $3\times10^{11}$  \\
 \hline
 B & $4\times 2_L\times 2_R~ [{\bf 126}]$ & $10^{13.8}$ & $5 \times10^{10}$  \\
\hline
C &  $3_C\times 2_L\times 2_R\times 1_{B-L}~ [{\bf 16}]$ & $10^{10.6}$ & $3 \times10^6$  \\
 \hline
 C &$3_C\times 2_L\times 2_R \times 1_{B-L} ~[{\bf 126}]$ & $10^{8.6}$ & $6 \times 10^3$  \\
\hline
\end{tabular}
\label{tab:schemes}
\end{table}

Finally, we must specify the identity of the NETDM candidate in the context described above.
The DM can be in the ${\bf 126}$ or ${\bf 144}$ representations of SO(10). 
There are several mechanisms to render the DM mass light \cite{Frigerio:2009wf}, one of which  is through a fine-tuning of the SO(10) couplings contributing with different Clebsh-Gordan coefficients (see for example, \cite{Slansky:1981yr} and \cite{Fukuyama:2004ps}) to the masses of the various ${\bf 126}$ components. For example, for the group ${\cal G}_A$:
\be
\overline{{\bf 126}}(M+ y_{45}{\bf 45}_H + y_{210}{\bf 210}_H){\bf 126}
\label{mft}
\ee
where $M\sim M_{GUT}$, and a ${\cal G}_A$ singlet in a linear combination of  ${\bf 210}_H$ and  ${\bf 45}_H$ has a vev at the GUT scale. $m_\chi$ is then given by a linear combination of $M$ and the vev and can be tuned to small values, while all other particles inside the {\bf126} live close to $M_{GUT}$.

\section{Discussion}

Unfortunately, the chance of detection (direct or indirect) of NETDM with a massive mediator $Z'$   is nearly hopeless. Indeed, the diagram
for the direct detection process, measuring the elastic scattering off a nucleus, proceeds
through the $t-$channel exchange of the $Z'$ boson, and is proportional to $1/M^4_{Z'}$ 
yielding a negligible cross-section. 
In addition, due to the present low velocity of dark matter in our galaxy 
($\simeq 200 ~ \mathrm{km/s}$), the indirect detection prospects from 
$s-$channel $Z'$ annihilation $\chi \chi \rightarrow Z' \rightarrow ff$ proportional
to $s^2/M_{Z'}^4$ is also negligible.

As we have seen in Eq.\ref{Eq:solution}, the production of dark matter
occurs in the very early Universe at the epoch of reheating. 
A similar mechanism 
(though fundamentally completely different) where a dark matter candidate is 
produced close to the reheating time is the case of the gravitino \cite{EHNOS,Gravitino}. Indeed,
in both cases equilibrium is never reached and the relic abundance is
produced from the thermal background to attain the decoupling value $\Gamma/H$, with $H$ the Hubble constant
and $\Gamma = \langle \sigma v \rangle n_f$ the production rate. However, in the case of SO(10), 
the cross section decreases with the temperature like 
$\langle \sigma v \rangle_{Z'} \propto T^2/ M_{Z'}^4$, whereas in the case of the gravitino
 the cross section is constant $\langle \sigma v \rangle_{3/2} \propto 1/M^2_P$ implying
 $Y(T) \propto T_{RH}$.

Finally, we note that cases B and C (in Table I) 
predict reheating temperatures which are  larger (B) or smaller (C) than the case under
consideration. Case A would also be compatible 
with successful thermal leptogenesis with a zero initial state abundance of right--handed 
neutrino \cite{leptogenesis1, leptogenesis2}. 
However in the cases B and C, the persistence of the SU(2)$_R$ symmetry would
imply that the cancelation in Eq. \ref{mft} would leave behind a light  SU(2)$_R$ triplet (for DM inside a {\bf 126}) or doublet (for DM inside a {\bf 144}).
These would affect somewhat the beta functions for the RGE's but more importantly
leave behind a test of the model. In the triplet (doublet) case, we would expect three (at least two) nearly degenerate states: one with with charge 0, being the DM candidate, and also states with electric charge $\pm 1$ and $\pm 2$ (or $\pm 1$ in the doublet case).

\section*{Conclusion}

In this work, we have shown that it is possible to produce dark matter through non--equilibrium thermal processes in 
the context of $SO(10)$ models which respect the WMAP constraints. Insisting on gauge
coupling unification, we have  demonstrated that there exists a tight link between the reheating
temperature and the scheme of the $SO(10)$ breaking to the SM gauge group. Interestingly,
the numerical values we obtained are quite high and very compatible with inflationary and leptogenesis-like models.

\noindent {\bf Acknowledgements. }  The authors would like  to thank M. Tytgat,
T. Hambye, E. Dudas, M.Voloshin,  E. Fernandez-Martinez, K. Tarachenko
 and J.M. Moreno for very useful discussions.
This  work was supported by
the French ANR TAPDMS {\bf ANR-09-JCJC-0146}  and the Spanish MICINN's
Consolider-Ingenio 2010 Programme  under grant  Multi-Dark {\bf CSD2009-00064}. B.Z. acknowledges 
Consolider-Ingenio PAU CSD2007-00060, CPAN CSD2007-00042, under  the contract {\bf FPA2010-17747}. 
Y.M.  acknowledges partial support from the European Union FP7 ITN INVISIBLES (Marie
Curie Actions, PITN- GA-2011- 289442). The work of K.A.O. was supported in part
by DOE grant DE--FG02--94ER--40823 at the University of Minnesota.

\end{document}